# Using Bibliometrics to Detect Unconventional Authorship Practices and Examine Their Impact on Global Research Metrics, 2019-2023

Lokman I. Meho[*] and Elie A. Akl

Faculty of Medicine, American University of Beirut, Beirut, Lebanon

**Abstract:** Between 2019 and 2023, sixteen universities increased their research output by over fifteen times the global average, alongside significant changes in authorship dynamics (e.g., decreased first authorship, rise in hyperprolific authors, increased multi-affiliations, and increased authors per publication rate). Using bibliometric methods, this study detected patterns suggesting a reliance on unconventional authorship practices, such as gift, honorary, and sold authorship, to inflate publication metrics. The study underscores the need for reforms by universities, policymakers, funding agencies, ranking agencies, accreditation bodies, scholarly publishers, and researchers to maintain academic integrity and ensure the reliability of global ranking systems.

## Introduction

Universities face increasing pressure to improve their standings in global ranking systems, given the significant influence of these rankings on students, faculty, funding, and policy decisions (Hazelkorn, 2015; Hazelkorn et al., 2021; Rovito et al., 2021). Many ranking systems emphasize research output as a key measure of institutional success, prompting universities to intensify efforts to increase their publication and citation numbers (Rhein et al., 2023; Sheeja, 2018; Vernon et al., 2018). This pursuit has sometimes led universities to adopt unconventional authorship practices, such as hiring productive external authors for a specific period to boost perceived research output and impact. However, these practices compromise academic integrity and distort ranking systems (Bhattacharjee, 2011; Biagioli et al., 2019; Moosa, 2024; Pachter, 2014; Trung, 2020). While not illegal, they exemplify attempts to artificially enhance rankings, regardless of their limitations (Moosa, 2024).

Recent dramatic increases in publication output at some institutions, far exceeding the global average, raise concerns about practices such as hyper, guest, gift, honorary, sold, and ghost authorship (Teixeira da Silva et al., 2016) (see definitions in Box 1). Hosseini et al. (2022) additionally suggest a mutual influence between ethical issues of authorship and increases in the average number of authors per publication.

> **Box 1: Definitions.** Hyperprolific authorship involves individuals who publish multiple papers per month over prolonged periods, straining the scientific community's understanding of normative standards of authorship and credit (Ioannidis et al., 2018; Pendlebury, 2024). Guest authorship involves the mutual inclusion of names among researchers to inflate publication counts, while gift authorship includes mentors, colleagues, or department heads as a sign of gratitude despite minimal contribution

---

[*] To whom all correspondence should be made: lmeho@aub.edu.lb

> (Ali, 2021; Ioannidis et al., 2024; Morreim et al., 2023). Honorary authorship adds senior researchers' names to gain favor, often at the expense of junior researchers (Kwee et al., 2023; Meursinge Reynders et al., 2024). Sold authorship involves paying for authorship slots in publications, and ghost authorship omits actual contributors (Chirico et al., 2023; Gureyev et al., 2022).

All these practices undermine research integrity and skew the evaluation of scholarly output. As a result, they have emerged as a critical issue (Chirico et al., 2023; Gureyev et al., 2022). Despite advancements and stringent ethical standards, current detection methods remain insufficient and largely focus on individual acceptable or unacceptable behavior (Resnik et al., 2015, 2016; Teixeira da Silva, 2023; Teixeira et al., 2016). There is a pressing need for more robust tools to uncover, at the institutional or aggregate level, unconventional authorship practices and their impact on global research metrics. Bibliometric approaches show significant promise in addressing these challenges (Biagioli et al., 2019; Gureev et al., 2019; Gureyev et al., 2022; Ioannidis et al., 2024).

Bibliometric evaluations can enhance the integrity of the research process and offer hope for a more transparent and fairer academic environment. These evaluations identify deviations from expected patterns in scholarly output, using authorship patterns, citation analysis, and other metrics to flag potential misconduct (Agarwal et al., 2016). By identifying these flags through bibliometric analyses, we can promote fair authorship practices and maintain the integrity of academic publishing and university ranking systems. As an example of a similar effort, in its 2023 edition of Highly Cited Researchers, Clarivate excluded more than 500 candidates due to extreme hyper-authorship, excessive self-citation, and unusual collaborative group citation activity patterns (Catanzaro, 2024; Pendlebury, 2024).

This study examines a group of universities that have recently exhibited extraordinary publication growth and changes in authorship patterns. The study aims to provide insights into the prevalence and impact of potential practices such as hyper, gift, and sold authorship on the evaluation of scholarly output.

**Materials and Methods**

We first identified the 1,000 most productive universities in 2023 via the Scopus database. For each of these universities, we collected bibliometric data for the period from January 2019 to December 2023 using various sources, including Scopus, Web of Science, Essential Science Indicators, InCites, and SciVal (Hazelkorn et al., 2021). We then focused on the 50 universities with the highest rate of increase in research output during this period. These universities experienced publication growth exceeding 130%—a figure over fifteen times the global average of 8.7% among academic institutions (InCites).

Recognizing that significant increases in research output can stem from various factors— such as governmental/institutional pressure or incentives (Abritis et al., 2017; Chou et al., 2016; Li, 2016; Paruzel-Czachura et al., 2021), mergers or expansions (Han et al., 2022; Handoyo et al., 2024), increased research funding (Ebadi & Schiffauerova, 2016; Mao et al., 2024), recruitment and retention of highly productive faculty (Rovito et al., 2021), enhanced international collaboration and partnerships (Marginson, 2022), shifts in publishing practices (Guskov, 2018), the growth of doctoral and postdoctoral programs (Sarrico, 2022), and publishing in predatory journals (Guskov, 2018)—as suggested by Gureyev et al. (2022) and Ioannidis et al. (2024), we concentrated on universities exhibiting substantial deviations from expected patterns in scholarly activity, more specifically, on those university displaying substantial drops in first authorship rates and significant increases in international collaboration rates over a short timeframe, as these traits may suggest the use of unconventional authorship practices referred to earlier. We focused on first authorship and international collaboration rates because they are readily available in research analytics tools like InCites and SciVal.

Accordingly, we identified a cohort of 16 universities ranked among the top 20 globally for the most significant decreases in first authorship rates or the largest increases in international collaboration rates from 2019 to 2023. This Study Group comprises nine public universities from Saudi Arabia and the United Arab Emirates and seven private universities from Egypt, India, Iraq, and Lebanon. Overall, eleven universities are less than 30 years old, while five have been established for over 50 years. A PRISMA-type chart in Box 2 shows a summary of the selection process for identifying the Study Group universities (Page et al., 2021).

> **Box 2:** Summary of the selection process for identifying the Study Group universities
>
> **Universities identified via Scopus (1,000 most published universities in 2023)**
> (n=1,000)
> Excluded: 9,427 institutions (3,000 of which published more than 100 articles each)
>
> **Universities with bibliometric data collected**
> (n=1,000)
>
> **Universities with >130% publication growth from 2019 to 2023 (or 15 times the world average among academic institutions)**
> (n=50)
> Excluded: 950 universities
>
> **Universities with substantial publication growth and:**
> **(1) ranked among the world's top 20 in drops in first authorship rates and/or**
> **(2) ranked among the world's top 20 in increases in international collaboration rates**
> (n=16)
> Excluded: 34 universities
>
> **Final study group: 16 universities**

The 34 excluded universities include 23 from China, seven from Saudi Arabia, and one from Morocco, Pakistan, the United Arab Emirates, and Vietnam.

For comparative analysis, we formed a Control Group of seven universities consistently ranked among the top 200 in the Academic Ranking of World Universities (Shanghai Ranking) to represent institutions that adhere to conventional authorship and affiliation practices. We excluded universities with medical schools (e.g., Australian National University, Brown University, Case Western, University of Basel, and the University of Bern) from the Control Group to ensure a more balanced comparison (Table 1).

We compared the two groups for the increase in publications overall and growth in the number of subject-specific publications, as well as trends in first authorship rates, number of authors per article, hyper-authorship, multiple affiliations, and international collaboration. We used only those publications referred to as "articles" or "reviews" (articles hereafter). Drawing on data from Scopus and employing SciVal and a methodology akin to Halevi et al. (2023), we tallied each author's publication count from each university and analyzed their institutional and country affiliations. To assess the robustness of our findings, we conducted sensitivity analyses and thoroughly cross-referenced our data using the mentioned databases and tools.

Ioannidis et al. (2018) defined hyperprolific authors as individuals who published over 72 articles, conference papers, substantive comments, or reviews in a year and found that 70% of them admitted not fulfilling the four ICMJE criteria more than 25% of the time. For this study, we used a threshold of 36 articles per year to better highlight differences among institutions. We define 'external authors' as individuals who are engaged in sold authorship, listing universities as secondary affiliations in over half of their publications associated with that institution (see Chirico et al., 2023; Guskov et al., 2018) or include individuals from other institutions as co-authors who do not meet standard authorship criteria (i.e., 'gift' authorship) (Gülen et al., 2020; Teixeira da Silva et al., 2016). Multi-affiliation is a phenomenon where authors of a scientific article have multiple affiliations, often from different institutions or countries (Hottenrott et al., 2022). According to Halevi et al. (2023), the past decade has seen a substantial rise in the number of affiliations listed by authors in scientific papers, with some listing over 20 institutions, raising concerns about the genuine contributions of these authors at each institution they claim to be affiliated with (Gök et al., 2024).

**Results**

*Surge in Overall Research Output*

Figure 1 highlights the contrasting trajectories of research output between the Study and Control groups from 2019 to 2023. The sixteen universities in the Study Group demonstrated substantial increases in publication output, ranging from 166% to 1474%, resulting in an overall increase of 266% (or over 30 times the world average). In contrast, the Control Group universities exhibited only modest changes in publication output, with variations ranging from an 8% decline to a 59% increase, leading to an overall increase of 10% during the same period (slightly over the 8.7% world average).

This significant surge in research output propelled the Study Group's global publication rankings dramatically upward. According to SciVal, their median rank improved from over 2000th in 2019 to 575th in 2023. Notably, four institutions ascended into the ranks of the world's top 300 most published universities. Conversely, the Control Group experienced a notable decline in their research output rankings, slipping from a median world rank of 275th to 369th, with four institutions dropping out of the top 300 globally.

### *Inflation in Discipline-Specific Research*

The dramatic increase in research output among the sixteen Study Group universities is further underscored when analyzed by discipline. In 2019, only one of these universities ranked among the world's top 100 most-published institutions in any of the 21 subject categories tracked by Clarivate's Essential Science Indicators. By 2023, however, six universities had broken into the top 100 a total of thirty-four times across thirteen subject categories, with significant representation in science, technology, engineering, and mathematics fields. In contrast, the Control Group declined in several areas, reducing its representation among the world's top 100 most-published universities from 17 appearances in nine subject categories in 2019 to 13 appearances in eight categories in 2023 (Table 2). It is important to note that apart from King Saud University and the three universities in India, virtually none of the other Study Group members have doctoral programs in the subject categories where they now rank among the 100 most published universities in the world.

Expanding the analysis to include the 300 most published universities globally in each subject category, the Study Group's presence becomes even more pronounced. Moving from 17 representations in 2019 by two universities to fourteen universities appearing 78 times across eighteen subject categories in 2023, the Study Group showed substantial gains in fields such as chemistry, computer science, engineering, environment & and ecology, materials science, mathematics, pharmacology and toxicology, and physics, with five or more universities in each category. Shifts in research output rankings were less pronounced in the medical, health, and social sciences. Conversely, the Control Group saw a decrease in representation among the world's 300 most published universities by field, dropping from 60 appearances in 2019 to 44 in 2023.

### *Decline in First Author Publications*

The percentage of university publications listing their researchers as first authors reflects the balance between internal and external contributions to research projects (Kharasch et al., 2021; Marušić et al., 2018). According to InCites, the global average for first authorship among academic institutions has slightly declined, from 53% in 2019 to 50% in 2023. Disciplinary averages for 2023 ranged from 42% in clinical medicine to 47-50% in the life, physical, and social sciences and 52-56% in computer science, engineering, and mathematics.

For the Study Group, there was a pronounced decline in first authorship publications, dropping from an average of 50% in 2019 to 28% in 2023. This decline is more than seven times the global average decrease. Notably, eleven universities within this group experienced the largest margins of decline globally. As a result, thirteen of the sixteen Study Group universities are now ranked among the world's 20 institutions with the lowest rates of first authorship, falling from an overall median rank of 713th to 992nd among the world's 1,000 most published universities in 2023 (Table 3).

Conversely, the Control Group exhibited modest declines in first authorship, decreasing from an average of 48% in 2019 to 43% in 2023. The median rank for this group remained relatively stable, shifting slightly from 632nd in 2019 to 634th in 2023.

### *Rise in Hyperprolific Authorship*

Based on SciVal data, we found a substantial rise in hyperprolific authors within the Study Group compared to the Control Group. The number of hyperprolific authors in the Study Group increased dramatically, from 18 (approximately one per institution) in 2019 to 260 (approximately 16 per institution) in 2023. Conversely, the Control Group maintained the same number of hyper-prolific authors in 2019 compared to 2023, with an average of two per institution each year (Table 4).

The publication record of an engineering faculty member from one of the Study Group members serves as a noteworthy example of these hyperprolific authors, potentially resulting from gift, honorary, or sold authorship (see also Abalkina, 2023; Chirico et al., 2023; Gülen et al., 2020). Scopus data reveal that from 2004 to 2021, this faculty averaged one article per year. Remarkably, their output jumped to 231 articles in 2022 and escalated to 516 articles in 2023, averaging over 1.4 articles per day and representing 38% of the entire university's research output that year. With 80% of articles published in top-quartile journals, the faculty was the first author on three articles and the sole corresponding author on 48 (or 6%) of the 747 articles in which they were listed on the byline in 2022-2023. This pattern suggests a limited personal contribution, potentially indicative of nominal authorship where the substantive research and writing might be handled by external authors (in this case, predominantly from Pakistan, China, and India).

Another illustrative example is a faculty member listing multiple Study Group institutions as their affiliation. Initially averaging five articles per year from 2016 to 2020, this faculty member's output surged to an average of 95 articles annually from 2021 to 2023. Notably, they list multiple affiliations in 83% of their publications (with an overall average of 3.4 affiliations per article), often including two to six international affiliations. As indicated in the next section, this practice is prevalent across the Study Group institutions.

### *Increase in Multi-Affiliated Publications*

When comparing the Study Group with the Control Group, significant differences in the trends of multiple affiliations from 2019 to 2023 become apparent. In 2019, the proportion of the Study Group articles where the institutions' authors listed multiple affiliations was 18% (excluding papers where co-authors listed the same institution as the sole affiliation). Although this proportion remained unchanged by 2023, the variation among individual institutions was notable. Three institutions increased their proportion by over 28 percentage points during this period. Overall, in 2023, five members of the Study Group institutions exhibited rates of listing multiple affiliations three to tenfold higher than the Control Group, which remained stable at 8% during the five years (Table 5).

### *Growth in Authorship Per Publication*

In the Study Group, the increase in the number of authors per article was particularly pronounced, rising by 28% from 5.0 in 2019 to 6.4 in 2023. In contrast, the Control Group saw a smaller increase of 17%, with the average number of authors per article growing from 6.9 to 8.1 over the same period. As a reference, and according to InCites, the global average number of authors per article increased by 7%, from 3.6 authors per article in 2019 to 3.9 in 2023.

### *Trends in Collaborative Publication Patterns*

In the Study Group, the percentage of articles that overlapped among two or more institutions was 6% in 2019. However, by 2023, this figure had increased by twelve percentage points to 18%. Among authors who contributed more than ten publications to the Study Group, we identified 50 individuals who credited their work to two or more Study Group institutions, indicating a widespread practice of network-driven authorship, which could be even more extensive in a larger consortium of institutions or across a broader network of universities. Conversely, the Control Group exhibited a more modest increase in collaboration between one institution and the other. The percentage of articles involving multiple Control Group institutions grew from 2% in 2019 to 3% in 2023.

### *Surge in International Collaboration*

For the Study Group, there was a pronounced increase in the proportion of articles with international collaboration, increasing from an average of 70% in 2019 to 81% in 2023. This increase is nearly three times the global average increase, which stands at less than 4% during the same period. Notably, six universities within this group experienced the largest margins of increase globally. As a result, half of the sixteen universities are now ranked among the world's 20 institutions with the highest rates of international collaboration, rising from a median rank of 68th in 2019 to 20th among the world's 1,000 most published universities in 2023.

Conversely, the Control Group exhibited modest increases in international collaboration, increasing from an average of 59% in 2019 to 62% in 2023. The median rank for this group remained relatively stable, shifting slightly from 392nd in 2019 to 410th in 2023 (Table 6).

VOSviewer-generated maps further illuminate the transition within the Study Group (van Eck et al., 2010). Figure 2 shows the institutional co-authorship network at the beginning of our study period in 2019, highlighting a network involving only 27 external institutions collaborating with any member of the Study Group on over 90 articles (the minimum number of articles published by a Group member). In contrast, Figure 3 from 2023 reveals a dramatic expansion, with the number of collaborating institutions skyrocketing to 254, representing an 840% increase. This map illustrates a vastly more connected and seemingly collaborative environment, though the validity of these collaborations is questionable, given the high proportion of secondary affiliations, the relatively disproportionate increases in the number of authors per paper, and the unusual collaborative group publication activity patterns.

For comparative context, Figures 4 and 5 depict the institutional co-authorship network maps for the Control Group in 2019 and 2023, respectively. These universities maintained a more stable collaboration landscape, with a moderate increase in collaborating institutions (from 137 in 2019 to 175 in 2023—a 28% increase).

**Discussion**

This study used bibliometric methods to detect unconventional authorship practices and examine their impact on global research metrics. Our findings reveal significant and concerning trends in authorship practices at a subset of universities from 2019 to 2023. The dramatic increase in research output among these institutions, far surpassing global averages, coupled with notable changes in authorship dynamics, suggests a potential reliance on gift, honorary, and sold authorship practices. These practices have significant implications for academic integrity and the reliability of global research metrics.

Specifically, we investigated the publication practices of sixteen universities that exhibited extraordinary growth in research output alongside changes in authorship dynamics during this period. Compared to a Control Group of established universities, the Study Group displayed considerable surges in publication output, often accompanied by several notable trends.

- **Reduced First Authorship:** The study group exhibited a pronounced decline in first authorship publications, which may indicate an increased reliance on external contributions. This trend is particularly noteworthy as first authorship is often viewed as a marker of significant contribution to the research.

- **Rise in Hyperprolific Authors:** The Study Group saw a dramatic increase in the number of individuals credited with an unusually high volume of publications within a short period. This hyperprolific authorship raises questions about the sustainability and authenticity of these contributions, as maintaining such high output levels is generally challenging without resorting to unconventional practices (Tóth et al., 2024).

- **Increased Multi-Affiliations:** The percentage of publications in which authors listed affiliations with multiple institutions rose significantly. This trend, which often involves authors affiliating with institutions in different countries (Halevi et al., 2023;

Hottenrott et al., 2022), raises concerns about these authors' genuine contributions to each institution's research output.

- **Inflated Authorship:** The Study Group also showed a significant rise in the average number of authors per publication compared to the Control Group. This increase may not always reflect true intellectual contributions from all listed authors (Hosseini et al., 2022).

- **Unusual Collaborative Patterns:** A substantial increase in co-authored publications between institutions within the Study Group was identified, often exceeding the level of collaboration normally expected based on research focus. This is largely due to relying on the same authors to boost publication metrics.

- **Explosive Growth in International Research Collaboration (IRC):** The Study Group experienced a dramatic rise in the proportion of internationally co-authored publications, far exceeding the global average increase. While IRC is generally beneficial, such a rapid increase suggests a possible overreliance on international partners to enhance research output figures. Previous studies have raised concerns about using international co-authorship as a reliable proxy for IRC measurement due to the variability in its dynamics and underlying motivations (Chen et al., 2019; Katz et al., 1997; Kuan et al., 2024; Luukkonen et al., 1993; Marginson, 2022). Our findings align with the arguments made by Gök and Karaulova (2023), who suggest that unconventional forms of IRC, such as IRC via multiple affiliations, are becoming more prevalent and must be scrutinized to ensure accurate assessments of research collaboration. They found that approximately 20% of co-publications are labeled international solely because of an author's second affiliation, highlighting the need for a nuanced understanding of IRC (Hottenrott et al., 2021).

The findings suggest that unconventional authorship practices, such as hyperprolific authorship, listing of an excessive number of affiliations, and guest, gift, and sold authorship, may be contributing factors to the observed surge in publication metrics. These practices, while not inherently illegal, raise concerns about the integrity of academic publishing and the reliability of university ranking systems. By identifying institutions with unusual publication patterns, we hope to foster a dialogue on maintaining high ethical standards in research.

Our findings highlight the potential of bibliometric evaluations in identifying deviations from expected patterns in scholarly output. By leveraging these evaluation methods, we can promote fair authorship practices and uphold the integrity of academic publishing.

The observed surges in research output within the Study Group may not reflect a genuine increase in home-grown research activity. The concerning trends in authorship dynamics suggest the need for further examination and dialogue among universities, policymakers, and ranking agencies. The universities in the Study Group appear to inflate their publication metrics by relying on external authors. This reliance on external collaborators to boost publication output suggests a shift from internally driven scholarly output towards possibly outsourced work that minimally involves the institution's faculty members and researchers in primary roles. The example of a faculty member with an abrupt increase in publication output and another faculty listing an unreasonable number of international affiliations exemplify potential nominal authorship practices, where actual research contributions may be minimal.

To foster a robust research culture that is sustainable and impactful, universities need to nurture their talented researchers and cultivate a vibrant, collaborative environment without overreliance on external contributors. Current trends in global ranking manipulations underscore the need for international bodies and academic institutions to critically reassess and refine ranking methodologies to ensure they reflect true academic excellence and integrity.

*Unsustainable Research Growth*

An example of the potential consequences of inflated publication rates is observed in the case of Taif University. According to Scopus, the institution peaked in publications in 2022 with 4,690 articles, up from 513 in 2019, only to see a notable drop to 2,377 in 2023. This drastic fluctuation suggests that the rapid increases seen across some universities will probably not be sustainable in the long term, potentially leading to a loss of trust in their published research. This instability exemplifies the risks associated with aggressive publication strategies to boost rankings rather than foster genuine academic progress. Moreover, Taif University also experienced a substantial reduction in the number of hyperprolific authors, dropping from 37 in 2022 to 14 in 2023, further illustrating the transient nature of such publication strategies. Similar trends are observed in the cases of Future University in Egypt and Al-Mustaqbal University, where during January-June 2024, they published only 30% as many articles as in 2023, compared to a world average of 51%.

Parallels have been observed in other institutions in the past (Alhuthali et al., 2022; Ansede, April 18, 2023; Bhattacharjee, 2011; Catanzaro, 2023; Catanzaro, 2024; SIRSI Academic, 2023). For example, Vietnam's Duy Tan University published 481 articles in 2018, peaking at 2,668 in 2020, but following public scrutiny (Trung, 2020), its output dipped to 846 in 2022 and 860 in 2023. Similarly, Ton Duc Thang University saw its publications increase from 302 in 2016 to 3,339 in 2020, and after similar public exposure, numbers fell to 755 in 2022 and 554 in 2023. Both universities dropped out of the Shanghai Ranking in 2023 after reaching a peak of 601-700 in 2021.

These cases underscore a challenge where universities may use practices that artificially inflate their research output to enhance their standings in global rankings. This trend not only misrepresents their actual academic contributions but also questions the sustainability and authenticity of their research advancements. Such phenomena, echoing the pitfalls seen at Taif University and other institutions, suggest that initial surges in publication numbers may often be driven by unsustainable practices that do not reflect genuine scholarly activity.

*The Urgency for Reform: A Call to Action*

The reliance on networks of external authors and unconventional authorship practices to increase research outputs might threaten academic research's integrity and undermine public trust in academic institutions. This issue also directly biases the outcomes of ranking systems, compromising their reliability and usefulness. To address this systemic challenge effectively, a concerted effort is required from all stakeholders in the academic community. Below are expanded proposed measures:

- **Universities:** Establish more stringent guidelines and policies for granting secondary affiliations and allowing primary faculty to hold such affiliations. Enhance internal review mechanisms to detect and address suspect authorship patterns. Endorse and communicate institutional values that reflect integrity and ethical conduct. Moreover, universities should make more effort to educate their faculty about appropriate and inappropriate authorship in scientific publications (Ali, 2021; Alshogran et al., 2018).

- **Policymakers:** Develop and enforce regulations that address questionable publication practices and provide clear guidelines for ethical authorship and collaboration (see, for example, Ansede, April 2, 2023). Encourage transparency in reporting research activities and outputs through legislative and administrative measures.

- **Funding Agencies:** Initiate audits to explore the potential misuse of external authors in funded projects and by applicants seeking funding.

- **Ranking Agencies:** Reconsider how to account for the number of publications and citations concerning primary versus secondary affiliations. Develop red flag indicators like those discussed in this paper and incorporate them into evaluation frameworks. Also, qualitative assessments of research relevance and genuine international collaboration should be incorporated. Consider focusing on metrics that incentivize homegrown research.

- **Accreditation Agencies:** Establish review guidelines and initiate audits to detect unconventional or unethical practices to boost publication records. Enforce strict compliance with the relevant accreditation standards related to the university's mission and values, faculty recruitment and affiliation practices, and research productivity and integrity.

- **Scholarly Publishers:** Strengthen peer-review processes to identify inappropriate authorship and affiliation claims more effectively, especially for networks of hyperprolific co-authors with multiple affiliations. Require detailed justifications for multiple secondary affiliations.

- **Researchers:** Maintain rigorous ethical standards in authorship and promote transparency within their academic environments. Actively report any suspicions of authorship fraud. Utilize resources like the Center for Scientific Integrity (Retraction Watch) for guidelines on responsible research conduct, including authorship ethics. The Committee on Publication Ethics (COPE) also provides guidelines for responsible authorship.

The challenges the Control Group universities encounter underscore the need for comprehensive reforms and stricter guidelines. Addressing the issues highlighted in this study is crucial for maintaining a trustworthy and authentic academic environment.

Implementing the aforementioned measures, coupled with a renewed commitment to conventional publication practices, is essential for preserving the integrity of scholarly work. Without intervention, the ongoing trends will continue, eroding trust in the academic record, obstructing true scholarly collaboration, and diverting resources at institutions from truly advancing academic excellence.

**Limitations**

This study analyzes unconventional authorship practices among sixteen universities that have exhibited extraordinary growth in research output, reduced first authorship, rise in hyperprolific authors, increased multi-affiliations, and increased authorship counts. However, several limitations should be noted:

**Sample Size and Scope:** This study focused on a cohort of sixteen universities ranked among the world's 1,000 most published universities in 2023. While this has provided valuable insights, it may not capture the full spectrum of unconventional authorship practices globally. Future studies could expand the sample size by including more universities from diverse regions and varying levels of research output.

**Temporal Scope:** This study examined data from 2019 to 2023. While this period captures recent trends, it may not fully reflect longer-term patterns in authorship practices. Future studies could benefit from extending the temporal scope to include data from earlier years to identify historical trends and the evolution of these practices over time.

**Institutional Contexts:** The study included universities from specific countries and regions. Differences in institutional policies, research cultures, and external pressures across various regions may influence authorship practices. Future research should aim to understand how these factors may vary globally.

**Interpretation of Collaborative Patterns:** While this study identified unusual collaborative patterns, interpreting these patterns accurately remains challenging. Future research could employ qualitative methods, such as case studies or interviews, to gain deeper insights into the motivations and mechanisms behind these collaborations.

These limitations highlight areas for future research that can build upon and extend our findings, providing a more comprehensive understanding of unconventional authorship practices and their impact on global research metrics.

**Funding:** None

**Author contributions:**
   Conceptualization: LIM, EAA
   Methodology: LIM, EAA
   Investigation: S LIM, EAA
   Visualization: LIM
   Supervision: LIM, EAA
   Writing—original draft: LIM
   Writing—review & editing: LIM, EAA

**Competing interests:** The authors declare that they are affiliated with a university that is a peer institution to one of the universities included in the Study Group.

**Data and materials availability:** All data are in the text, tables, and charts.

**Table 1. Percentage of change in publication counts for universities in the study and control groups between 2019 and 2023**

| University name (year founded) | Country | Type | Size | # of articles | | Change | World rank in # of articles | |
|---|---|---|---|---|---|---|---|---|
| | | | | **2019** | **2023** | | **2019** | **2023** |
| **Study Group** | | | | | | | | |
| Future University in Egypt (2006) | Egypt | Private | M | 127 | 1,373 | 981% | 2000+ | 986 |
| Chandigarh University (2012) | India | Private | L | 362 | 2,327 | 543% | 2000+ | 583 |
| GLA University (2010) | India | Private | M | 259 | 1,572 | 507% | 2000+ | 870 |
| Lovely Professional University | India | Private | L | 838 | 2,302 | 175% | 1152 | 593 |
| University of Petroleum and Energy Studies (2003) | India | Private | M | 308 | 1,569 | 411% | 2000+ | 871 |
| Al-Mustaqbal University (2010) | Iraq | Private | M | 91 | 1,432 | 1,474% | 2000+ | 947 |
| Lebanese American University (1924) | Lebanon | Private | M | 315 | 2,637 | 737% | 2000+ | 500 |
| Al-Imam Mohammad Ibn Saud Islamic University (1974) | Saudi Arabia | Public | XL | 364 | 1,588 | 336% | 2000+ | 865 |
| King Khalid University (1998) | Saudi Arabia | Public | L | 1,327 | 5,158 | 289% | 778 | 199 |
| King Saud University | Saudi Arabia | | XL | 4,490 | 11,962 | 166% | 175 | 29 |
| Prince Sattam Bin Abdulaziz University (2009) | Saudi Arabia | Public | L | 750 | 4,388 | 485% | 1254 | 255 |
| Princess Nourah Bint Abdulrahman University (1970) | Saudi Arabia | Public | L | 471 | 4,468 | 849% | 1749 | 250 |
| Taif University (2004) | Saudi Arabia | Public | XL | 513 | 2,377 | 363% | 1662 | 567 |
| Umm Al-Qura University (1950) | Saudi Arabia | Public | XL | 589 | 3,072 | 422% | 1508 | 419 |
| University of Tabuk | Saudi Arabia | Public | L | 414 | 1,389 | 236% | 2000+ | 969 |
| University of Sharjah (1997) | United Arab Emirates | Public | M | 758 | 2,465 | 225% | 1238 | 541 |
| **Control Group** | | | | | | | | |
| City University of Hong Kong (1994) | Hong Kong | Public | M | 3,450 | 5,481 | 59% | 275 | 181 |
| Swiss Federal Institute of Technology Lausanne (1969) | Switzerland | Public | M | 3,495 | 3,472 | -1% | 269 | 369 |
| California Institute of Technology | United States | Private | S | 3,634 | 3,719 | 2% | 258 | 341 |
| Carnegie Mellon University (1900) | United States | Private | M | 2,266 | 2,299 | 1% | 465 | 589 |
| Princeton University | United States | Private | M | 3,595 | 3,819 | 6% | 266 | 325 |
| Rice University (1912) | United States | Private | M | 1,765 | 1,851 | 5% | 605 | 747 |
| University of California, Santa Barbara (1891) | United States | Public | M | 3,263 | 2,999 | -8% | 299 | 430 |

**Data sources:** SciVal for publications and QS and Times Higher Education for size (June 2024).

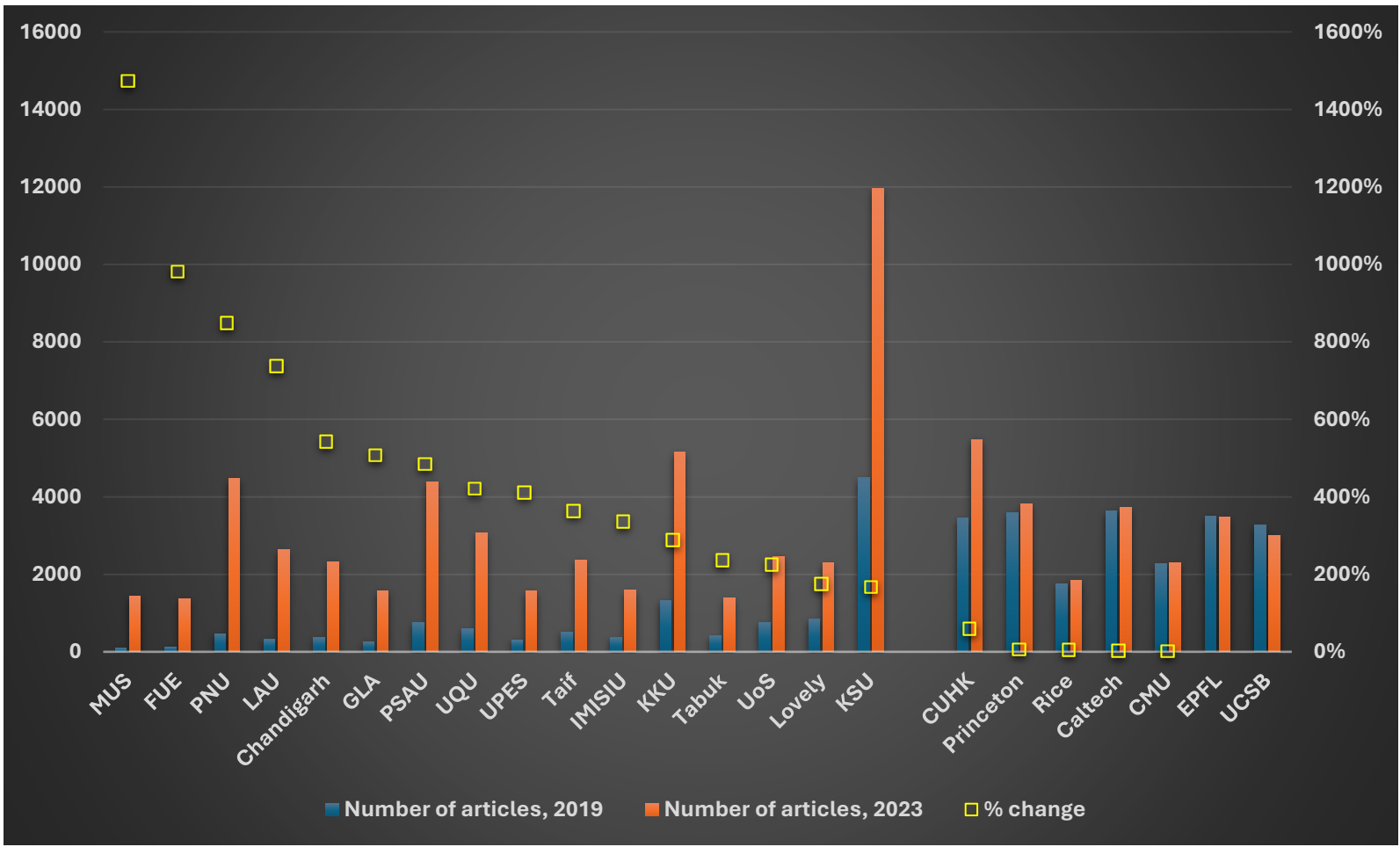

**Figure 1. Percentage change in publications counts for universities in the study and control groups between 2019 and 2023.** A blank space separates the Study Group (left) from the Control Group (right). Chandigarh=Chandigarh University (India), CMU=Carnegie Mellon University (USA), CUHK=City University of Hong Kong (Hong Kong), EPFL=Swiss Federal Institute of Technology Lausanne (Switzerland), FUE=Future University in Egypt (Egypt), GLA=GLA University (India), IMISIU= Al-Imam Mohammad Ibn Saud Islamic University (Saudi Arabia), KKU=King Khalid University (Saudi Arabia), KSU=King Saud University (Saudi Arabia), LAU=Lebanese American University (Lebanon), Lovely=Lovely Professional University (India), MUS=Al-Mustaqbal University (Iraq), PNU=Princess Nourah Bint Abdulrahman University (Saudi Arabia), PSAU=Prince Sattam Bin Abdulaziz University (Saudi Arabia), Tabuk=University of Tabuk (Saudi Arabia), Taif=Taif University (Saudi Arabia), UCSB=University of California Santa Barbara (United States), UoS=University of Sharjah (United Arab Emirates), UPES=University of Petroleum and Energy Studies (India), UQU=Umm Al-Qura University (Saudi Arabia). The percentage of change for EPFL and UCSB does not show because they are negative 1% and 8%, respectively. **Data Source:** SciVal (June 2024).

**Table 2. Change in the number of articles and world rank for specific universities by subject category between 2019 and 2023 (top 100 in each category)**

| Name | # of articles, 2019 | # of articles, 2023 | World rank in # of articles, 2019 | World rank in # of articles, 2023 |
|---|---|---|---|---|
| **Agricultural Sciences (n=1)** | | | | |
| **Study Group** | | | | |
| King Saud University | 188 | 510 | **72** | **12** |
| **Control Group** | | | | |
| None ranked among the world's 100 most published | | | | |
| **Biology & Biochemistry (n=1)** | | | | |
| **Study Group** | | | | |
| King Saud University | 321 | 611 | **71** | **7** |
| **Control Group** | | | | |
| None ranked among the world's 100 most published | | | | |
| **Chemistry (n=3)** | | | | |
| **Study Group** | | | | |
| King Saud University | 603 | 2266 | **58** | **2** |
| King Khalid University | 200 | 957 | **446** | **33** |
| Princess Nourah Bint Abdulrahman University | 61 | 658 | **1266** | **60** |
| **Control Group** | | | | |
| Swiss Federal Institute of Technology Lausanne | 554 | 368 | **77** | **Dropped out** |
| **Computer Science (n=5)** | | | | |
| **Study Group** | | | | |
| King Saud University | 198 | 297 | **55** | **53** |
| Lebanese American University | 21 | 289 | **1053** | **57** |
| Prince Sattam Bin Abdulaziz University | 20 | 278 | **1085** | **61** |
| Princess Nourah Bint Abdulrahman University | 13 | 266 | **1466** | **67** |
| King Khalid University | 32 | 202 | **752** | **96** |
| **Control Group** | | | | |
| City University of Hong Kong | 316 | 449 | **23** | **30** |
| Carnegie Mellon University | 195 | 178 | **56** | **Dropped out** |
| **Economics & Business (n=1)** | | | | |
| **Study Group** | | | | |
| Lebanese American University | 26 | 235 | **731** | **24** |
| **Control Group** | | | | |
| City University of Hong Kong | 150 | 143 | **63** | **86** |
| **Engineering (n=4)** | | | | |
| **Study Group** | | | | |
| King Saud University | 459 | 1487 | **131** | **38** |
| King Khalid University | 171 | 988 | **469** | **58** |
| Prince Sattam Bin Abdulaziz University | 98 | 918 | **775** | **66** |
| Princess Nourah Bint Abdulrahman University | 23 | 706 | **1994** | **100** |
| **Control Group** | | | | |
| City University of Hong Kong | 845 | 1200 | **45** | **50** |
| **Environment/Ecology (n=1)** | | | | |
| **Study Group** | | | | |
| King Saud University | 184 | 946 | **195** | **4** |
| **Control Group** | | | | |
| None ranked among the world's 100 most published | | | | |
| **Geosciences (n=0)** | | | | |
| **Study Group** | | | | |
| None ranked among the world's 100 most published | | | | |
| **Control Group** | | | | |
| California Institute of Technology | 721 | 576 | **7** | **21** |
| Princeton University | 257 | 231 | **89** | **100** |
| **Materials Science (n=2)** | | | | |
| **Study Group** | | | | |
| King Saud University | 374 | 1006 | **111** | **33** |
| King Khalid University | 189 | 598 | **275** | **71** |
| **Control Group** | | | | |
| City University of Hong Kong | 669 | 1005 | **45** | **34** |
| Swiss Federal Institute of Technology Lausanne | 407 | 337 | **95** | **Dropped out** |
| **Mathematics (n=6)** | | | | |
| **Study Group** | | | | |
| King Saud University | 151 | 470 | **94** | **2** |
| Princess Nourah Bint Abdulrahman University | 18 | 328 | **1195** | **7** |
| Prince Sattam Bin Abdulaziz University | 46 | 297 | **578** | **13** |
| King Khalid University | 51 | 252 | **504** | **18** |
| Umm Al Qura University | 18 | 190 | **1195** | **46** |
| Lebanese American University | 2 | 149 | **2825** | **98** |
| **Control Group** | | | | |
| Princeton University | 214 | 184 | **30** | **50** |
| **Microbiology (n=1)** | | | | |
| **Study Group** | | | | |
| King Saud University | 55 | 148 | **192** | **37** |
| **Control Group** | | | | |
| None ranked among the world's 100 most published | | | | |
| **Pharmacology & Toxicology (n=5)** | | | | |
| **Study Group** | | | | |
| King Saud University | 263 | 685 | **28** | **1** |
| Prince Sattam Bin Abdulaziz University | 73 | 284 | **310** | **37** |
| King Khalid University | 76 | 226 | **292** | **64** |
| Umm Al Qura University | 41 | 210 | **583** | **70** |
| Princess Nourah Bint Abdulrahman University | 26 | 183 | **822** | **89** |
| **Control Group** | | | | |
| None ranked among the world's 100 most published | | | | |
| **Physics (n=3)** | | | | |
| **Study Group** | | | | |
| King Saud University | 168 | 416 | **283** | **66** |
| King Khalid University | 165 | 367 | **294** | **78** |
| Princess Nourah Bint Abdulrahman University | 17 | 338 | **1661** | **90** |
| **Control Group** | | | | |
| Princeton University | 655 | **555** | 29 | 32 |
| California Institute of Technology | 482 | 436 | **55** | **60** |
| Swiss Federal Institute of Technology Lausanne | 462 | 446 | **59** | **54** |
| University of California Santa Barbara | 435 | 319 | **59** | **97** |
| **Plant & Animal Science (n=1)** | | | | |
| **Study Group** | | | | |
| King Saud University | 225 | 703 | **132** | **10** |
| **Control Group** | | | | |
| None ranked among the world's 100 most published | | | | |
| **Space Science (n=0)** | | | | |
| **Study Group** | | | | |
| None ranked among the world's 100 most published | | | | |
| **Control Group** | | | | |
| California Institute of Technology | 956 | 1030 | **1** | **1** |
| Princeton University | 423 | 526 | **16** | **14** |
| University of California, Santa Barbara | 156 | 141 | **85** | **Dropped out** |

**Data source:** Essential Science Indicators, via InCites (June 2024).

**Table 3. Decline in the proportion of and world rank in first authorships for universities in the study and control groups between 2019 and 2023**

| Institution | % first author, 2019 | % first author, 2023 | World rank in % articles as first author, 2019 | World rank in % articles as first author, 2023 |
|---|---|---|---|---|
| **Study Group** | | | | |
| Al-Mustaqbal University College | 35 | 11 | 998 | 999 |
| Chandigarh University | 60 | 26 | 308 | 994 |
| Future University in Egypt | 44 | 10 | 879 | 1000 |
| GLA University | 72 | 35 | 77 | 977 |
| Imam Mohammad Ibn Saud Islamic University (IMSIU) | 51 | 34 | 677 | 984 |
| King Khalid University | 48 | 18 | 807 | 997 |
| King Saud University | 48 | 29 | 815 | 990 |
| Lebanese American University | 55 | 17 | 448 | 998 |
| Lovely Professional University | 65 | 50 | 197 | 495 |
| Prince Sattam Bin Abdulaziz University | 50 | 28 | 749 | 992 |
| Princess Nourah Bint Abdulrahman University | 44 | 27 | 953 | 993 |
| Taif University | 55 | 24 | 461 | 996 |
| Umm Al Qura University | 47 | 33 | 887 | 986 |
| University of Petroleum & Energy Studies (UPES) | 60 | 32 | 301 | 987 |
| University of Sharjah | 49 | 35 | 770 | 975 |
| University of Tabuk | 52 | 35 | 604 | 982 |
| **Control Group** | | | | |
| California Institute of Technology | 43 | 37 | 972 | 965 |
| Carnegie Mellon University | 51 | 47 | 632 | 634 |
| City University of Hong Kong | 45 | 39 | 932 | 939 |
| Ecole Polytechnique Federale de Lausanne | 55 | 50 | 452 | 516 |
| Princeton University | 52 | 48 | 599 | 583 |
| Rice University | 51 | 44 | 650 | 783 |
| University of California Santa Barbara | 52 | 50 | 598 | 512 |

**Data source:** InCites (June 2024).

**Table 4. Changes in the count of hyper-prolific authors for universities in the study and control groups between 2019 and 2023**

| Institution | 2019 | 2020 | 2021 | 2022 | 2023 |
|---|---|---|---|---|---|
| **Study Group** | | | | | |
| King Saud University | 9 | 44 | 53 | 47 | 89 |
| King Khalid University | 7 | 10 | 31 | 49 | 26 |
| Lebanese American University | 0 | 0 | 1 | 2 | 26 |
| Prince Sattam Bin Abdulaziz University | 1 | 4 | 15 | 40 | 23 |
| Princess Nourah Bint Abdulrahman University | 0 | 1 | 3 | 10 | 22 |
| Taif University | 0 | 1 | 25 | 37 | 14 |
| University of Sharjah | 0 | 1 | 8 | 9 | 11 |
| Umm Al-Qura University | 0 | 0 | 2 | 9 | 10 |
| GLA University | 0 | 0 | 0 | 3 | 8 |
| University of Petroleum and Energy Studies | 0 | 0 | 0 | 4 | 7 |
| Chandigarh University | 0 | 0 | 0 | 2 | 6 |
| Future University in Egypt | 0 | 0 | 0 | 5 | 6 |
| Al-Mustaqbal University College | 0 | 0 | 0 | 5 | 5 |
| Lovely Professional University | 0 | 0 | 1 | 3 | 5 |
| Al-Imam Mohammad Ibn Saud Islamic University | 0 | 0 | 0 | 0 | 2 |
| University of Tabuk | 1 | 2 | 1 | 3 | 0 |
| **Control Group** | | | | | |
| City University of Hong Kong | 6 | 9 | 9 | 10 | 10 |
| Princeton University | 2 | 4 | 3 | 2 | 2 |
| California Institute of Technology | 2 | 5 | 3 | 3 | 1 |
| Swiss Federal Institute of Technology Lausanne | 3 | 3 | 3 | 1 | 0 |
| University of California at Santa Barbara | 2 | 4 | 1 | 0 | 1 |
| Carnegie Mellon University | 1 | 1 | 1 | 1 | 1 |
| Rice University | 1 | 1 | 1 | 1 | 1 |

**Data Source:** SciVal (June 2024).

**Table 5. Trends in listing multiple affiliations by authors of universities** in the study and control groups from 2019 to 2023*

| Institution | % listing multiple affiliations, 2019 | % listing multiple affiliations, 2023 | Change (percentage points) |
|---|---|---|---|
| **Study Group** | | | |
| Lebanese American University | 15% | 76% | 61% |
| Chandigarh University | 9% | 39% | 31% |
| University of Petroleum and Energy Studies | 5% | 34% | 29% |
| University of Sharjah | 19% | 27% | 8% |
| Prince Sattam Bin Abdulaziz University | 29% | 24% | -5% |
| Al-Imam Mohammad Ibn Saud Islamic University | 28% | 21% | -7% |
| University of Tabuk | 27% | 13% | -14% |
| King Khalid University | 21% | 12% | -9% |
| Umm Al-Qura University | 31% | 10% | -21% |
| Lovely Professional University | 2% | 10% | 8% |
| Taif University | 43% | 9% | -34% |
| Future University in Egypt | 26% | 7% | -19% |
| Al-Mustaqbal University College | 21% | 5% | -16% |
| Princess Nourah Bint Abdulrahman University | 21% | 4% | -17% |
| GLA University | 4% | 4% | 0% |
| King Saud University | 11% | 3% | -8% |
| **Control Group** | | | |
| City University of Hong Kong | 18% | 22% | 4% |
| Princeton University | 8% | 8% | 0% |
| Swiss Federal Institute of Technology Lausanne | 8% | 8% | -1% |
| California Institute of Technology | 7% | 6% | -1% |
| Rice University | 7% | 5% | -2% |
| University of California, Santa Barbara | 6% | 5% | -1% |
| Carnegie Mellon University | 5% | 4% | -1% |

*Excludes papers where co-authors listed the same institution as the sole affiliation. **Data Source:** Scopus (June 2024).

**Table 6. Trends in the proportion of articles with international collaboration in the study and control groups from 2019 to 2023***

| Name | % International Collaborations (2019) | % International Collaborations (2023) | World rank in % of articles with int'l collab. (2019) | World rank in % of articles with int'l collab. (2023) |
|---|---|---|---|---|
| **Study Group** | | | | |
| Lebanese American University | 54 | 95 | 250 | 1 |
| King Khalid University | 75 | 89 | 15 | 3 |
| Princess Nourah bint Abdulrahman University | 72 | 89 | 24 | 4 |
| Future University in Egypt | 27 | 87 | 788 | 5 |
| University of Sharjah | 78 | 87 | 5 | 6 |
| Prince Sattam Bin Abdulaziz University | 74 | 85 | 17 | 7 |
| Taif University | 73 | 82 | 23 | 11 |
| Al-Mustaqbal University College | 48 | 79 | 372 | 18 |
| University of Tabuk | 76 | 78 | 13 | 21 |
| King Saud University | 74 | 78 | 18 | 22 |
| Imam Mohammad Ibn Saud Islamic University (IMSIU) | 61 | 77 | 109 | 23 |
| Umm Al Qura University | 72 | 77 | 27 | 28 |
| Chandigarh University | 33 | 66 | 643 | 118 |
| University of Petroleum & Energy Studies (UPES) | 17 | 64 | 839 | 141 |
| GLA University | 14 | 58 | 991 | 247 |
| Lovely Professional University | 33 | 52 | 653 | 333 |
| **Control Group** | | | | |
| Ecole Polytechnique Federale de Lausanne | 70 | 74 | 29 | 40 |
| California Institute of Technology | 59 | 60 | 140 | 219 |
| Princeton University | 49 | 52 | 345 | 344 |
| University of California Santa Barbara | 46 | 48 | 397 | 410 |
| Rice University | 46 | 47 | 392 | 426 |
| Carnegie Mellon University | 42 | 46 | 454 | 440 |
| City University of Hong Kong | 40 | 37 | 488 | 601 |

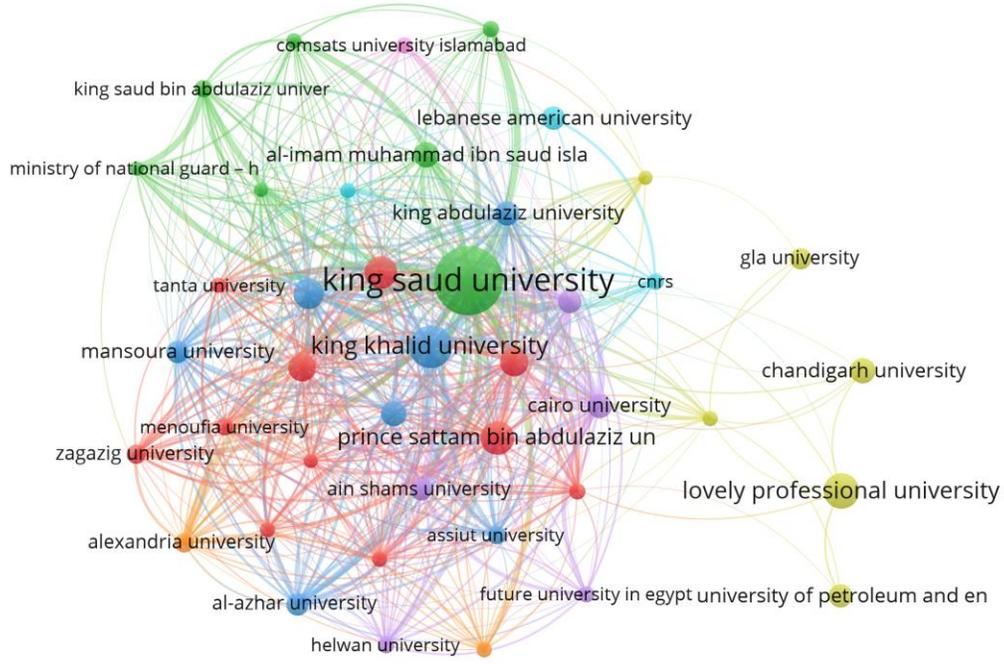
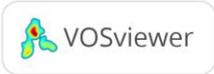

**Articles per institution: 91 | External institutions: 27 | Links: 547 | Total link strength: 7k | Clusters: 9**

**Figure 2. Institutional co-authorship network map of the Study Group in 2019** (i.e., at the beginning or before publication inflation). This map includes only institutions with over 90 articles (the minimum number of articles published by a member of the Study Group). The interactive version allows viewers to explore the connections and intensity of collaboration between the institutions. Total number of articles published by the Study Group in 2019 = 11,202.

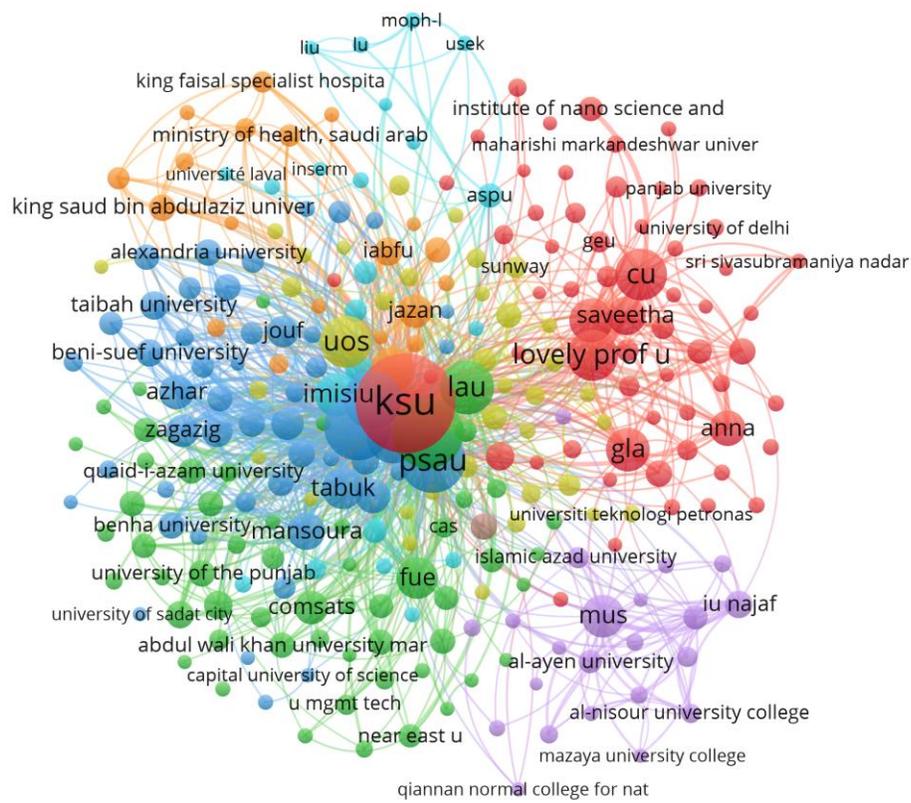
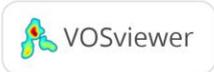

**Articles per institution: 91 | External institutions: 254 | Links: 15511 | Total link strength: 154k | Clusters: 8**

**Figure 3. Institutional co-authorship network map of the Study Group in 2023.** This map includes only institutions with over 90 articles. The interactive version allows viewers to explore the connections and intensity of collaboration between the institutions. Total number of articles published by the Study Group in 2023 = 41,026. Note the 840% increase in the number of institutions added to the network (from 27 in 2019 to 254 in 2023) among institutions that published over 90 articles in collaboration with the Study Group members.

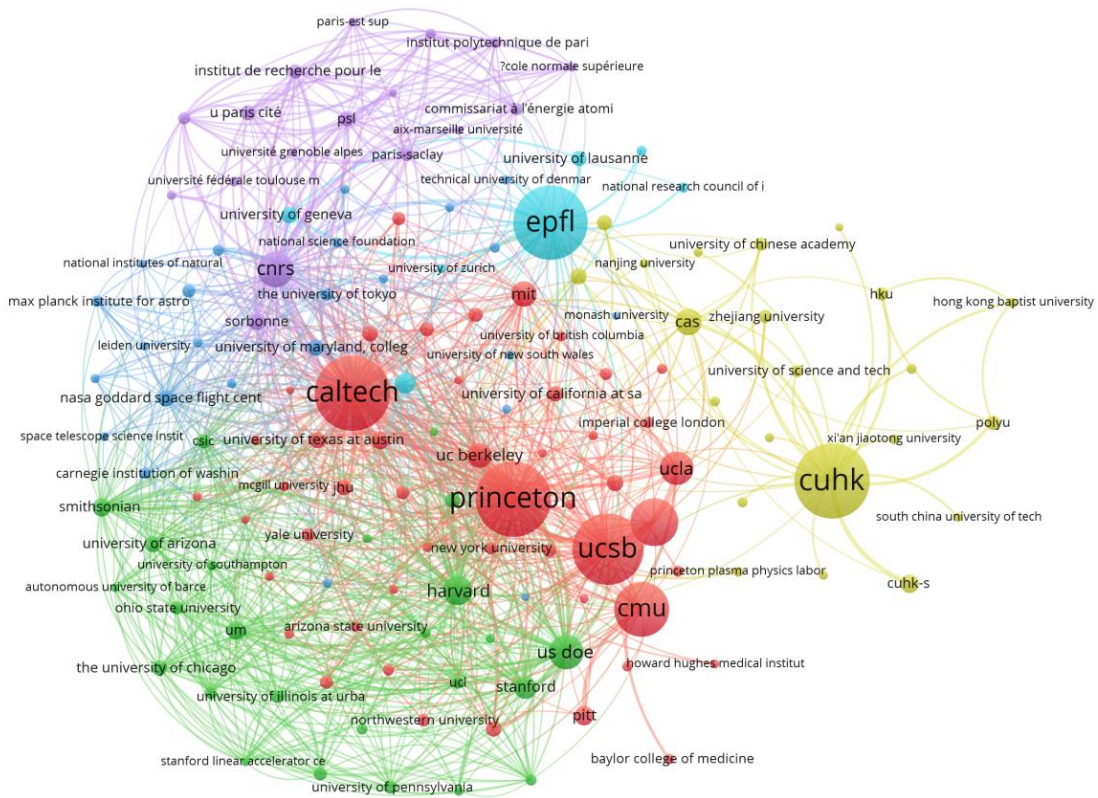
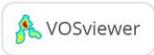

**Articles per institution: 91 | External institutions: 137 | Links: 8197| Total link strength: 94k | Clusters: 6**

**Figure 4. Institutional co-authorship network map of the Control Group in 2019**. This map includes only institutions with over 90 articles. The interactive version allows viewers to explore the connections and intensity of collaboration between the institutions. Total number of articles published by the Study Group in 2019 = 20,079.

**Articles per institution: 91 | External institutions: 175 | Links: 12757 | Total link strength: 158k | Clusters: 6**

**Figure 5. Institutional co-authorship network map of the Control Group in 2023**. This map includes only institutions with over 90 articles. The interactive version allows viewers to explore the connections and intensity of collaboration between the institutions. The total number of articles published by the Study Group in 2023 = 22,083. The group increased its network of institutions with over 90 articles in collaboration with each other by 28% (from 137 in 2019 to 175 institutions in 2023).